\documentclass[journal,twocolumn,11pt]{IEEEtran}
\usepackage[singlespacing,nodisplayskipstretch]{setspace}
\usepackage[utf8]{inputenc} 
\usepackage[T1]{fontenc}
\usepackage[colorlinks,pdfstartview=FitH,citecolor=blue,linkcolor=blue,urlcolor=blue]{hyperref}
\usepackage{url,comment}
\usepackage{ifthen}
\usepackage{cite}
\usepackage{braket}
\usepackage{graphicx}
\usepackage{amsmath,amssymb,amsthm}
\usepackage{cite,soul}
\usepackage{tikz}
\usepackage[capitalize]{cleveref}
\usepackage{wrapfig}
\usepackage{authblk}
\usepackage{flushend}

\usepackage{geometry}
\ifCLASSINFOpdf
\else
\fi

\usepackage[utf8]{inputenc}
\usepackage{enumitem}

\newcommand\ld[1]{{\color{black}#1}}

\title{QKD Based on Time-Entangled Photons and its Key-Rate Promise}
\author[1]{Lara Dolecek}
\author[2]{Emina Soljanin}
\affil[1]{ECE Department, University of California, Los Angeles, email: dolecek@ee.ucla.edu}
\affil[2]{ECE Department, Rutgers University, email: emina.soljanin@rutgers.edu }
\date{June 2022}
\setlist[itemize]{leftmargin=*}

\begin{document}

\maketitle

\begin{abstract}
 For secure practical systems, quantum key distribution (QKD) must provide high key rates over long distances. Time-entanglement-based QKD promises to increase the secret key rate and distribution distances compared to other QKD implementations. This article describes the major steps in QKD protocols, focusing on the nascent QKD technology based on high-dimensional time-bin entangled photons. We overview state-of-the-art from the information and coding theory perspective.
 In particular, we discuss the key rate loss due to single-photon detector imperfections.
 We hope the open questions posed and discussed in this paper will inspire information and coding theorists to contribute to and impact fledgling quantum applications and influence future quantum communication systems. 
\end{abstract}
\section{Introduction}

As predicted in the visionary works of Shannon, security is as essential to communication systems as reliability is. Not only did Shannon start the field of information theory with his groundbreaking work on the Mathematical Theory of Communication in 1948, but he also almost concurrently (in 1949) published another influential article, the one on information security \cite{shannon}. 
In this seminal work, Shannon proved that the one-time pad encryption scheme (known at least since the late 19th century) is information-theoretically secure. It achieves {\it perfect secrecy} where the a posteriori distribution of the encrypted messages is identical to the a priori distribution of the original.

Fast-forwarding decades of the digital revolution, Shannon's 
original ideas are as relevant as ever. We are now on the cusp of the next frontier in data communications, marked by quantum technologies. Quantum technologies have the potential to deliver unprecedented data security and processing that was previously unimaginable with classical systems. \ld{Not only has there been recent global interest and investment in quantum communication networks, but early implementations will only expand in the future \cite{accenture}.}
The future is arguably quantum, and we are at the dawn of this exciting new era.

 A vital component of any quantum communication system is Quantum Key Distribution (QKD). For widespread adoption of QKD - and truly secure quantum networks - it is imperative to provide high secret key rates over long distances. This article argues that QKD based on entangled photons could deliver on this challenge through concerted innovations in complementary domains: information and coding theory, and quantum physics.

State-of-the-art fiber-based QKD implementations have mostly reached the Mbits/s levels \cite{Diamanti_2016}, and some are even at the Gbps/s levels \cite{Shapiro}, over up to 400 km distances. While these results are encouraging, scaling to end-to-end implementations of wide-area networks decreases the practical key rates drastically due to various challenging implementation issues, including losses and noise, \cite{Diamanti_2016}. 
The inability to maximize the utility of information-bearing photons and to do so at low latency, has emerged as a bottleneck in practice, \cite{qinternet}.
 
 QKD protocols today operate under ``photon-starved'' conditions \cite{qinternet}. Despite an ongoing effort to improve these conditions with brighter sources and faster detectors, for the conventional discrete variable protocols, 
 the photons still carry less than a single bit through the channel. High-dimensional time entanglement is a promising strategy to break this bottleneck.

QKD schemes based on time-entangled photons extract key bits from the photon arrival times.
 Time-bin entanglement goes back to binary encoding schemes that distinguish only between early and late arrivals. 
 This information provides at most one bit of key per photon. In practical systems, the secret key rate is much lower because of noise and secrecy considerations. 
 Compared to the binary case, high-dimensional time-bin protocol promise to increase photon utilization by providing more precise measurement of the photon arrival times \cite{Lee:19}. However, practical photon detectors exhibit photon detection timing jitter, detector downtime, and dark current noise. 
 
 The adverse effects of various detector imperfections in QKD protocols have been recognized. A recent survey paper \cite{QKD:realistic} extensively studies secure quantum key distribution with realistic devices in the context of prepare-and-measure protocols, such as BB84. Key rates of time-entanglement-based QKD with detectors that exhibit timing jitter, downtime, and dark counts are addressed in \cite{QKD:BirnieCS22,Birnie:Thesis:2022,Cheng:Thesis:2022}.
 
 This short expository magazine article has a limited scope. For broader, deeper, and rigorous accounts, we refer the reader throughout the paper to specific technical papers and references therein. 
 Section~\ref{section:overview} describes the two fundamental steps of QKD protocols: raw key generation over a quantum channel and information reconciliation over the classical channel. 
 Section~\ref{section:time} focuses on time-entanglement QKD protocol raw key generation with time binning. It discusses the rate loss due to the non-ideal photon detectors. 
 
 In Section~\ref{section:coding}, we first explain that one-way information reconciliation can be naturally cast as a source coding problem with side information, for which forward error correction (FEC) schemes must be developed. We then devote that section to a brief survey of known results in FEC  for information reconciliation. Next, in Section~\ref{section:highdim}, we present challenges and progress to date on the multi-dimensional time-bin QKD focusing on the available FEC and modulation schemes.
 
  We conclude the article with a discussion and a collection of open problems of interest to (classical) information and coding theorists and quantum communication practitioners, presented in Section~\ref{section:discussion}.
This article does not attempt to cover all aspects of QKD-based systems; we refer interested readers to related surveys, e.g., \cite{Diamanti_2016}. 

\section{QKD Protocols}\label{section:overview}
\begin{figure*}[hbt]
    \centering
    \includegraphics{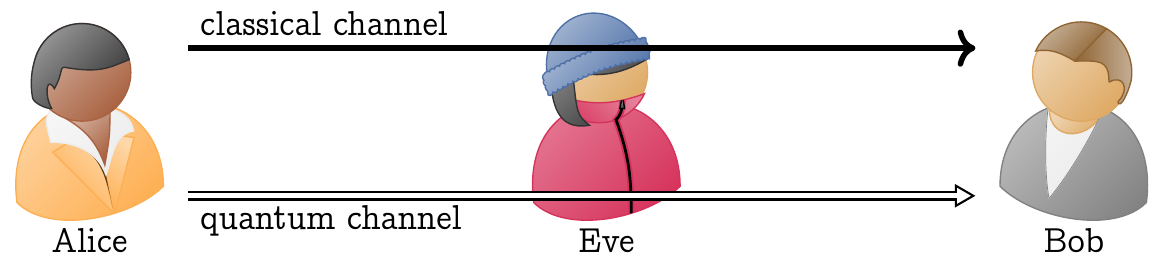}
    \caption{QKD: Alice and Bob establish a secret key by communicating over a quantum and a classical channel. Eve has access to both channels.}
    \label{fig:ProtocolClannels}
\end{figure*}

QKD refers to protocols that two parties, commonly referred to as Alice and Bob, can use to establish a secret key by communicating over a quantum and a classical channel that an eavesdropper, Eve, can access, as illustrated in Fig.~\ref{fig:ProtocolClannels}.
At the end of the protocol, Alice and Bob have identical uniformly random sequences known only to them, \ld{or they had aborted the protocol}. As described below, the QKD protocol requires communications through a quantum and a classical channel.

At a high level, there are two main QKD steps. Alice and Bob generate  \textit{raw key} bits in the first step using a quantum channel, which is essential for preventing undetected eavesdropping. 
At the end of this step, because of the noise in the system, Alice's and Bob's respective raw keys may disagree at some positions, be partly known to Eve, and may not be uniformly random. 

Alice and Bob process the raw key in the second step to establish a shared {\it secret key}. They communicate through the public classical channel to reconcile differences between their raw keys, amplify the privacy of the key concerning Eve's knowledge, and compress their sequences to achieve uniform randomness. 
They end up with a secret key that is private and hard to guess. 
After the classical postprocessing, the objective is to maximize the number of bits left in the secret key.

This paper focuses on QKD protocols that are based on time-entangled photons. In entanglement-based protocols, Alice, Bob, or some third party distribute entangled pairs of photons so that Alice and Bob each end up with one photon from each entangled pair. 		
The original protocol was proposed by Ekert in 1991 and is known as the E91 QKD \cite{QKD:ekert91}.
Suppose Alice and Bob share a set of $m$ entangled pairs of qubits in the state 
\begin{equation}
 \ket{\varphi}=\ket{00}+\ket{11})/\sqrt{2},
\label{eq:epr}   
\end{equation}
and Eve is not present. If they measure their respective states in 
the computational basis, they will get identical sequences of $m$ uniformly random bits, one for each photon pair. Thus, the scheme benefits from two properties of shared entanglement: randomness and correlation.  

Entangling two photons in polarization results in a two-qubit state in (\ref{eq:epr}). In such schemes, each entangled photon pair contributes one bit to the secret key at most. Entangling photons in time promises to give more bits per photon and thus improve photon utilization. 

\section{Time Entanglement QKD}\label{section:time}
We next describe the time-entanglement QKD scheme and the challenges in keeping its photon utilization promise. We refer the reader to \cite{QKD:BirnieCS22} for more detail.
\subsection{Generating Raw Key Bits}\label{subsection:rawkey}
\begin{figure}[hbt]
    \centering
    \includegraphics[scale=1.5]{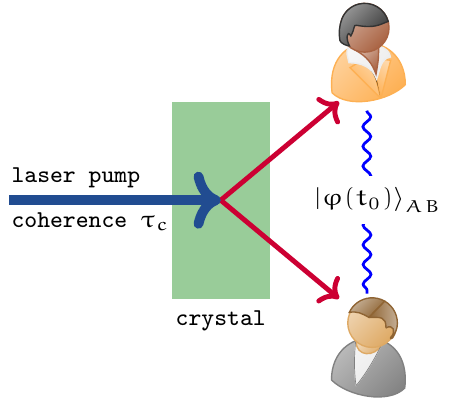}
    \caption{Generation of time-entangled photon pair in the states  $\displaystyle{\ket{\varphi(t_0)}_{AB}\propto\int_{t_0-\tau_c}^{t_0}\!\! dt~ \ket{t}_A\ket{t}_B}$. Ideally, Alice and Bob receive their individual photons at identical and uniformly
random times in the interval $[t_0,t_0-\tau_c$, but that is not true when they use practical, imperfect detectors.}
    \label{fig:source}
\end{figure}

In common time entanglement-based QKD schemes, an independent source (or one of the participants) randomly generates entangled photon pairs by Spontaneous Parametric Down Conversion (SPDC), as illustrated in Fig.~\ref{fig:source}.  
When we irradiate a non-linear crystal with a laser pump, SPDC may occur, resulting in two new entangled photons. The emission is equally likely to occur anywhere within a window equal to the pump coherence time $\tau_c$, and it does not depend on the previous emissions. 
Therefore, the entangled bi-photon state is given by
\[
\ket{\varphi(t_0)}_{AB}\propto\int_{t_0-\tau_c}^{t_0}\!\! dt~ \ket{t}_A\ket{t}_B,
\]
where one photon goes to Alice and the other to Bob.

Entangled photon inter-generation times are independent and identically exponentially distributed, giving a source of perfect randomness. 
Alice and Bob extract the raw key bits from the arrival times of entangled photons through time binning. Each individually discretizes their timeline into \textit{time bins} and groups them into \textit{time frames}. They record photon arrivals as occupied bins within frames. They then use the position of the photon bin within a time frame to generate random bits. The bit extraction follows pulse position modulation (PPM), first considered in~\cite{Kochman}. 
Under ideal conditions, photon inter-arrival times (as their inter-generation times) are independent and identically exponentially distributed. Several groups studied such systems, most notably \cite{QKD:Zhong15}, who constructed an end-to-end, high-dimensional time-entanglement experiment. 

\begin{figure}[hbt]
    \centering
    \includegraphics[scale=0.8]{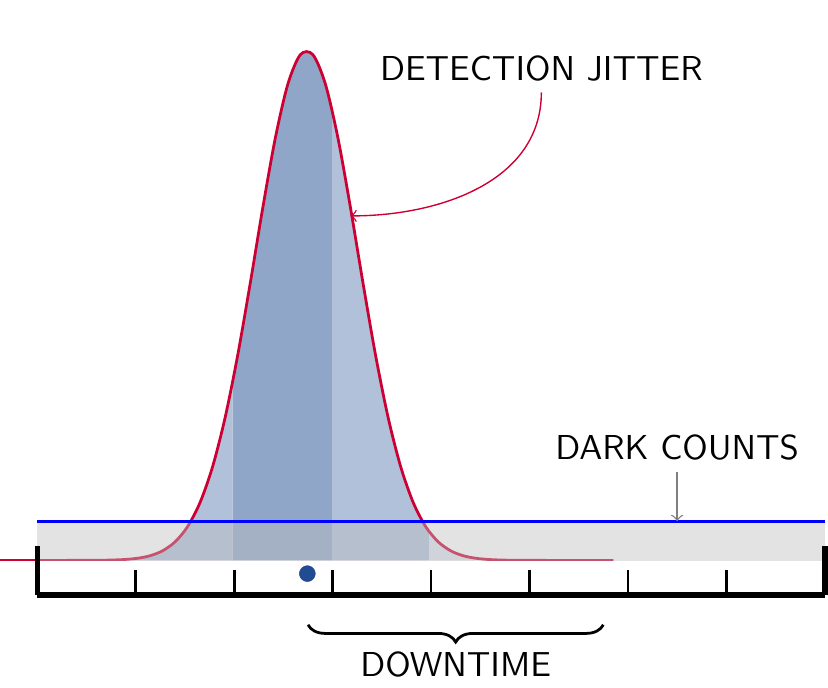}
    \caption{A frame with eight bins and a single occupied bin. A bin is identified with three bits. Detector imperfections include jitter, downtime, and dark counts. The Gaussian curve sketches the PDF of the detector jitter. Multiple bins may be affected, each with likelihood proportional to the area under the curve above it. Dark counts occur uniformly within the frame and here are represented by the uniform distribution.}
    \label{fig:jitteroverbins}
\end{figure}

 In PPM, Alice and Bob synchronize their clocks and discretize their timelines into fixed time frames, each consisting of $n$ time bins.
In PPM, Alice and Bob agree to retain only time frames in which precisely one bin is occupied while discarding all other frames. 
The maximum number of raw key bits that PPM decoding can extract from each retained frame is, therefore, $\log_2 n$. Figure~\ref{fig:jitteroverbins} shows a frame with eight bins, and indicates detection imperfections which we discuss later.
If Alice and Bob increase the number of bins per frame $n$, they will get more raw key bits per frame. However, their keys will potentially differ more because of the detection imperfections.

\subsection{An Illustrative Example}\label{section:example}

\begin{figure*}
    \centering
    \includegraphics{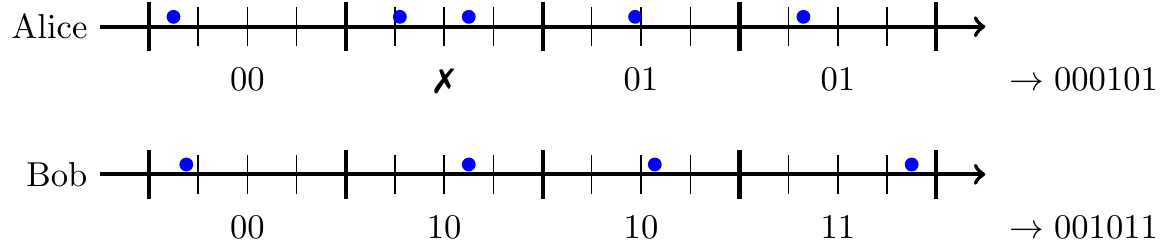}
    \caption{An example of arrival sequences at Alice (A) and Bob (B). Binary sequences on the right are what each party submits for information reconciliation.}
    \label{fig:example}
\end{figure*}

In Figure~\ref{fig:example}, we provide an example of binary sequences Alice and Bob have, illustrating how they agree and how they may potentially differ. 
We assume that Alice and Bob have each discretized their timeline into frames and divided each frame into four equally sized bins. Since there are four bins per frame, a bin contributes two bits of information, provided that it is the only occupied bin in the frame. Frames that have no occupied bins or more than one occupied bin are discarded; the former occurs due to photon losses and the latter due to dark counts.

In this example, we show four representative frames. In representing four bins in a frame by going left to right, we assign bit pairs as '00', '01', '10', and '11', in that order.
\begin{itemize} 
    \item In the leftmost frame, both Alice and Bob \ld{map their result to} '00' (the leftmost bin is occupied in both). Bit extraction is successful despite a slight arrival jitter.
    \item In the second frame, Alice detects two arrivals, with the spurious one due to dark counts. While Bob initially \ld{can map his result to} '01' for his second frame, upon receiving information from Alice that this frame is invalid, he too discards it.
    \item In the third frame, both Alice and Bob observe a single arrival. However, due to timing jitter, the two arrivals fall into adjacent bins. Alice \ld{maps her result to '01' whereas Bob maps his to} '10', resulting in a 2-bit discrepancy.
    \item In the rightmost frame, Alice and Bob each again detect a single arrival. In this example, two arrivals are due to dark counts and are uncorrelated (unbeknownst to Alice and Bob at this point). Alice \ld{maps her result to '01' and Bob his to} '11'. 
\end{itemize}

As illustrated by the examples, the effect of timing jitter, photon losses, and dark counts on the final key rate is dependent on the choice of bin width and size of a frame.

\subsection{Detector Imperfections}
The most common single-photon detectors are Superconducting Nanowire Single-Photon Detectors (SNSPDs).
Practical detectors suffer from jitter, dark counts, and downtime. 

Jitter errors occur because of imprecision in the time tagging, which causes discrepancies between Alice's and Bob's raw keys. Raw-key discrepancies reduce the secret key rate by increasing Alice and Bob's public exchange information rate for key reconciliation. Downtime is the time following a photon detection during which no other detection can occur. In their experiment, Zhong et al.\cite{QKD:Zhong15} used a 50:50 beam-splitter to distribute the photon arrivals to two detectors at each station to overcome the loss of photons caused by downtime. However, the detector downtime causes detection omissions and thus introduces memory into the system, altering the perceived arrivals such that they are no longer independent. This dependency alone reduces the secret key rate.
Dark counts are photons that do not come from the source of entangled photons and are primarily due to light leakage into the detector's optical lines. Dark counts arrive uniformly and independently at either detector. Since they are indistinguishable from the SPDC photons, they can cause significant errors if they make up a large fraction of the detected photons. 

We mathematically model these errors as follows; see also Figure~\ref{fig:jitteroverbins}. Let $U$ be the random variable modeling the arrival time of the entangled photons. Ideally, $U$ is uniformly distributed over the interval $[t_0-\tau_c,t_0]$, for some choice of time instance $t_0$, and where $\tau_c$ we recall is pump coherence time. However, the detector downtime introduces memory and makes $U$ non-uniform.
Because of the detector jitter, Alice's detector registers the  arrival at a time modeled by the random variable $T_A$, and Bob's detector registers the  arrival at a time modeled by the random variable $T_B$ where
\begin{equation}
 T_A = U +\eta_A ~\text{and} ~ T_B = U + \eta_B.   
\label{eq:co}
\end{equation}
Here $\eta_A$ and $\eta_B$ are independent zero-mean Gaussian random variables. 

Under the time-binning key extraction, Alice and Bob observe discrete correlated random variables identifying the occupied bin within the frame. When the number of bins per frame is $n$, Alice observes $X_n$ and Bob observes $Y_n$ given by
\begin{equation}
    X_n= U_n+J_{A,n}~\text{and}~Y_n= U_n+J_{B,n}
    \label{eq:dcm}
\end{equation}
where $U_n$ is uniform over $\{0,1,\dots,n-1\}$, and $J_{A,n}$ and $J_{B,n}$ have integer support, taking value $k$, $0 \leq k \leq n-1$, with the probability 
that depends on the noise statistics and the bin size.

\subsection{Eavesdropping Model}
 Since no passive eavesdropping is possible on a quantum channel,
 Alice and Bob can always detect the intercept-resend attack where Eve measures Alice's quantum states (photons) and then sends replacement states to Bob, prepared in the state she measures. They commonly pass a fraction of photons through a special interferometer to produce entangled photons in the maximally entangled state
$\ket{\varphi_{AB}}\propto\ket{0_A0_B}+\ket{1_A1_B}$. Alice and Bob can quantify Eve's information gain based on such photons by playing a variant of the CHSH game. We assume that they halt the key distribution if they detect an eavesdropper beyond the non-classical bound of the CHSH game. Thus, the raw keys that Alice and Bob receive have guaranteed security against eavesdropping on the quantum channel. 
Experimental systems implementing this kind of QKD protocol have been recently shown to achieve photon information efficiency up to 4.082 secure-key bits/photon, and a secure-key rate up to 237-kbit/s \ld{to provide security against collective attacks} \cite{QKD:wong21}. 

The intercept-resend attack is the simplest type of possible attacks, which are beyond the scope of this expository magazine article. Instead, we focus on the information that Eve gains by simply observing the communications over the public channel during the reconciliation phase we describe next. 

\subsection{Secret Key Rate}\label{section:secretkey}
The secret key rate is the ``maximum rate at which Alice and Bob can agree on a secret key while keeping the rate at which Eve obtains information arbitrarily small'', \cite{Maurer93a}. In the case of time-entanglement-based QKD, Alice and Bob obtain correlated streams of bits (raw keys) based on their measurements, as illustrated in Section~\ref{section:example}. However, they must communicate to agree on a key, i.e., reconcile their differences. Every communication required for this process must be considered \ld{public}, rendering popular Cascade protocol~\cite{cascade} inefficient; see also Section~\ref{section:overview}.

\begin{figure}[hbt]
    \centering
    \includegraphics{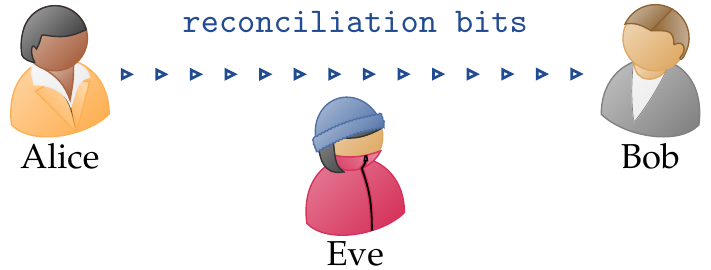}
    \caption{Alice sends bits to Bob over a public channel. Bob uses the bits to reconcile his raw key with Alice's.}
    \label{fig:reconciliation}
\end{figure}

Here, we consider one-way information reconciliation schemes in which Alice sends information about her sequence to Bob, who uses it to correct the differences between his and Alice's raw keys, as illustrated in Fig.~\ref{fig:reconciliation}. 

After one-way information reconciliation (to be discussed in the next section), Alice and Bob share Alice's initial raw key. However, since they communicated over a public channel, the shared key is not secret. To correct that, Alice and Bob perform privacy amplification. They 
compress their shared keys, establishing secrecy but shortening the key. 

\subsection{Rate Loss with Non-ideal Detectors}
\subsubsection{Rate Loss due to Detector Jitter} 

The secrecy capacity of our binning scheme for the number of bins equal to $n$ is $I(X_n;Y_n)$
where $X_n$ and $Y_n$ are given by \eqref{eq:dcm}. We say that $I(X,Y) = \lim_{n\to\infty} I(X_n;Y_n)$ is the ultimate achievable secret key rate (see, e.g., \cite[p.~567]{books:GK2011}).

We consider the following example.
Suppose that Alice's detector is noiseless and detects photons at the start of a bin. Bob's photon arrives (after) Alice's equally likely within time $\Delta$. When we split a frame of duration $T_f$ into $n$ bins,  the bin size is ${T_f}/n$.
Under these assumptions, we see that 
if  $n  \le T_f/\Delta$ (or equivalently $\Delta\le T_f/n$), Alice and Bob have identical raw key bits, and thus the secret key rate is
\[
 I(X_n;Y_n) = \log n.
 \]
On the other hand, when $n > T_f/\Delta$, we have  
		\begin{align*}
		    I(X_n;Y_n) = & H(X_n) - H(X_n|Y_n)\\
		    = &\log n -\log \Delta/({T_f}/n) =\log\frac{\Delta}{T_f}.
		\end{align*}
Therefore, increasing $n$ results in a secret key rate increase but only as long as $n  \le T_f/\Delta$.  

\subsubsection{Rate Loss due to Detector Downtime}
The non-zero downtime does not introduce errors. It introduces memory in Alice's and Bob's raw key bits. Thus Alice and Bob must compress their reconciled raw keys to achieve uniform randomness. The compression rate is a function of the detector downtime, photon generation rate, and the number of bins per frame. 

\begin{figure}[hbt]
    \centering
    \includegraphics[scale=0.75]{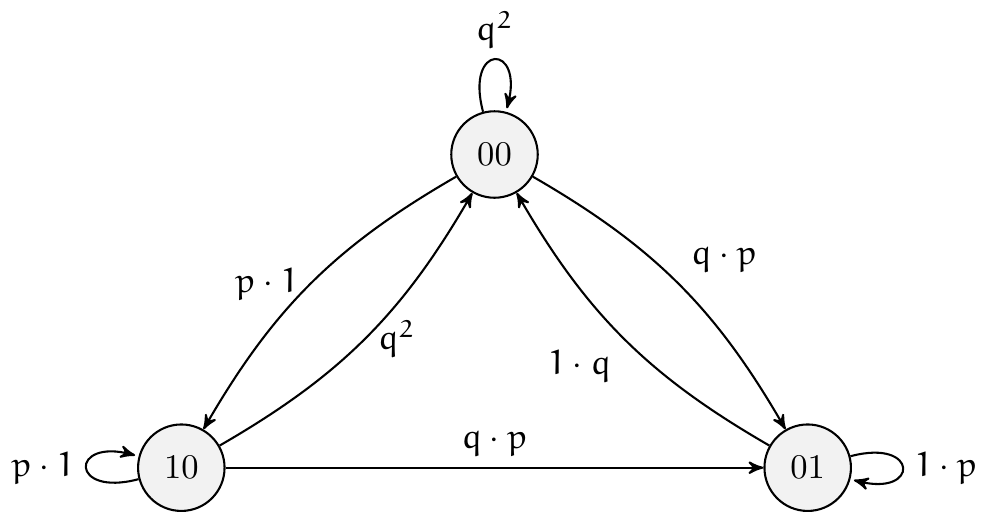}
    \caption{Markov chain modeling a system with two-bin frames (states of the chain) and detector downtime of one bin. The probability of photon arrival in a bin is $p$ and $q=1-p$. Because of the downtime, no frame will have two occupied bins.}
    \label{fig:MC}
\end{figure}

To characterize the impact of detector downtime on the system, we have to model combined detector and time binning operations by Markov Chains (MC). The entropy rate of the system's MC determines the minimum compression rate to guarantee the key's uniform randomness.
A simple example is shown in Fig.~\ref{fig:MC}.
These MCs can become very complex as the system parameters change. An algorithm to create them for various parameter values is presented in \cite{QKD:BirnieCS22} and implemented in an online tool available at \href{https://cc1539.github.io/qkd-binning-demo-2}{https://cc1539.github.io/qkd-binning-demo-2}.

\subsection{Time-Entanglement Rate Promise}
\begin{figure}
    \centering
    \includegraphics[scale=0.92]{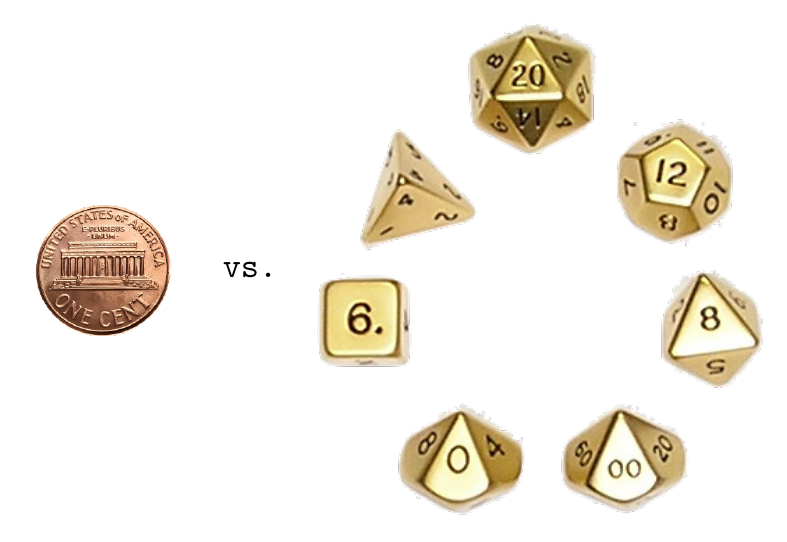}
    \caption{In principle, a system tossing a multi-faced coin can give more than a single bit of information per toss. However, if an icosahedron tossing system noise allows us to only distinguish between single and double-digit numbers, then that system acts as a penny-tossing one.}
    \label{fig:promise}
\end{figure}

Time-entanglement QKD promises to deliver more than one bit per photon as opposed to polarization-entanglement QKD, where each entangled photon pair contributes at most one bit to the secret key. To examine this promise in the light of the system description, it helps to consider tossing (multi-faced) coins, as illustrated in Fig.~\ref{fig:promise}.

The maximum key rate we can get with polarization entanglement corresponds to the information rate in a fair penny-tossing experiment, namely one bit. With time entanglement PPM, our coin becomes multi-faced, where the number of faces equals $n$, the number of bins in the frame. Thus the number of raw key bits with, e.g., the tetrahedron coin is $\log 4$, and with the octahedron is $\log 8$. However, increasing the number of coin sides also increases the effect of the system's imperfections. If Alice gets, e.g., four when she tosses her octahedron coin (i.e., measures the arrival time of her photon in an eight-bin frame), Bob may get four, three, or five on his side because of the detector jitter. The detector downtime (in the coin-tossing analogy) would make it impossible to observe small numbers after observing large numbers. 

For a given PPM frame duration (usually determined by the pump coherence time) and the detector noise parameters (jitter variance and downtime), $I(X_n; Y_n)$ can be maximized by selecting the number of bins per frame $n$; see \cite{QKD:BirnieCS22} for more detail. In principle, any rate $I(X_n; Y_n)$ can be achieved by appropriate error-correcting coding-based reconciliation schemes, which we discuss in the following sections. 

For additional improvements in photon utilization, we would have to have better equipment (sources of entangled photons and single-photon detectors) \cite{hyper:Chang21}. To increase photon utilization, adaptive PPM schemes have been proposed, see \cite{QKD:KarimiSW20} and references therein.

\section{FEC for QKD}\label{section:coding}

\subsection{Information-theoretic View}

Since Alice and Bob extract the secret key from correlated random variables, one-way information reconciliation amounts to Slepian-Wolf coding, a well-studied problem of source coding with side information. 
The general principle is as follows: Alice records her sequence of bits and sends the syndrome of that sequence to Bob. The syndrome is computed based on the parity check of a chosen code, and it is transmitted over the public classical channel. Bob uses his own sequence of bits along with the syndrome received from Alice to decode (reconcile) Alice's sequence. In Fig.~\ref{fig:reconciliation}, reconciliation bits are precisely syndrome bits generated by the chosen code at Alice's end.

The main challenge lies in mathematically constructing high-performance practical error recovery schemes compatible with practical physical systems capable of generating information-bearing photons while minimizing the information leakage to Eve. For such coding solutions to succeed in practice, they need high-performance and fast decoding algorithms.




Time-entanglement QKD described in Section~\ref{section:overview} represents an instance of a multi-dimensional (multi-bit) discrete variable (DV) QKD  since the information is represented as the index of one of the discrete-time bins. It is also possible for Alice to modulate coherent states, and for Bob to measure the amplitude and phase quadratures of the electromagnetic light field. This set up is referred to as continuous variable (CV) QKD. 
Since the results on FEC are arguably the most mature for the binary DV QKD, and since they serve as the necessary background for the research in FEC for high-dimensional time-entanglement, we devote the following subsection to the discussion of that literature.

\subsection{FEC for the canonical DV QKD models}


FEC  methods have been used with great success in many data transmission and storage applications. Graph-based codes such as low-density parity-check (LDPC) are among the most popular. LDPC codes have likewise been the primary mathematical technique in QKD information reconciliation to overcome
imperfect detectors and transmission noise.

For the binary DV QKD, initially motivated by encoding onto the polarization or phase of a photon, the early pioneering work \cite{boutros} keenly recognized that latency challenges associated with Cascade protocol could be overcome with appropriately designed channel codes; the focus of \cite{boutros} was to develop LDPC codes for the binary representation. In this scenario, Alice records a binary sequence $X$, and Bob records a binary sequence $Y$. Sequences $X$ and $Y$ are of the same length. 

Alice computes the syndrome of $X$ using the parity check matrix of a chosen LDPC code and transmits this syndrome to Bob over a public channel. Due to physical impairments that arise during entangled photon transmission, $X$ and $Y$ need not be the same. It was convenient to statistically relate $X$ to $Y$ through a binary symmetric channel (BSC) with cross-over probability $p$, where $p$ is assumed to be known by all parties. Based on the received syndrome and his own side information $Y$, Bob's task is to figure out what $X$ is. The work in \cite{boutros} observed that for the LDPC codes to be effective, they need to be optimized for the resultant BSC channel operating close to capacity. This was done by utilizing discrete density-evolution tools to optimize the degree distribution of LDPC codes and constructing resultant parity check matrices at various operating points (each resulting in a code with a different rate). The resultant codes are typically millions of bits in length to provide high performance. While these codes were individually highly optimized, they were, by design, structurally wholly separate. Such an approach entails extremely high implementation complexity, as each decoder, in principle, needs to be separately implemented at very long code lengths.


A collection of works subsequently focused on developing LDPC codes with better trade-offs than the fully interactive cascade and the baseline, single-pass LDPC coding. Binary rate-compatible LDPC codes had already been proposed in \cite{girod} in the context of distributed source coding and, as such, can also be appropriate for the QKD problem. Node puncturing and code shortening are conventional techniques that were successfully used in \cite{martin2010} to adapt the rate of a binary LDPC code for DV QKD (under the BSC channel).

Non-binary rate-compatible LDPC codes were proposed in \cite{Kasai2010InformationRF}. The advantage of these rate-compatible codes is that Alice can gradually reveal an increasing subset of her syndrome bits, which Bob uses in conjunction with his own side information to decode Alice's sequence. Code construction in~\cite{Kasai2010InformationRF} is based on starting with a ``mother code'' such that throughout the process, the same bipartite graph is used, thus at least partly alleviating the implementation challenges associated with the original LDPC proposals.

As in other (classical) applications, another promising channel coding technique is based on polar codes. Very long (order of a million bits) codes were again considered to achieve high efficiency. It was shown that the decoding latency based on polar codes was lower than that of comparable LDPC codes thanks to recursive decoding and that polar codes can be effective in both the DV and CV settings \cite{polar}.


We note at this point that this substantial body of work, while focusing on constructing codes possessing high efficiency along with rate compatibility and low complexity, uniformly assumes canonical and well-studied channel models, namely the BSC (for the DV setting) and AWGN (for the CV setting) channels, e.g., \cite{Kasai2010InformationRF, polar,kaist, milicevic2018}. 

It is apparent that while these assumptions can reap the benefits from vast literature on coding for BSC/AWGN channels in conventional communication applications, they fall short in capturing the intricate time-bin detector and channel impairments we previously discussed in Section~\ref{section:time}.

In the next section, we discuss recent progress on mathematical tools explicitly tailored for the high-dimensional time-entanglement QKD and show how such tools outperform existing methods built on canonical assumptions, thus closing the gap toward the promised key rate.

\section{Information Theory and Physics to the Rescue}\label{section:highdim}

Combining new ideas from quantum photonics on how to generate photon entanglement and new ideas from information theory on how to make the best use of such information-bearing photons will be necessary to unlock the full potential of QKD systems. 
Recall the description of how the raw key bits are generated in the time-bin protocol, cf.\ Section~\ref{subsection:rawkey}: arrival times of photons are stamped with a bin index within a frame.
In the baseline PPM, Alice and Bob keep only frames with a single photon for further processing, and all other frames (including empty frames and those with more than one stamped photon arrival) are discarded.  

While bits can be easily represented using a baseline PPM, as described in Section~\ref{subsection:rawkey}, well-designed adaptive modulation schemes can offer better performance. Work in \cite{QKD:Zhou13} investigated the performance of adaptive modulation schemes that do not necessarily discard frames with more than one detected photon. The paper analyzed and evaluated the performances of both fixed and adaptive PPM schemes based on a more realistic model that considers photon transmission losses, detection losses, and multi-pair events. Numerical results confirmed the significant benefits of the adaptive scheme. 

In a  recent work~\cite{QKD:KarimiSW20}, a more refined photon arrival was modeled, based on which the baseline PPM was first analyzed. The work ~\cite{QKD:KarimiSW20} demonstrated that this baseline scheme generates a significantly lower information rate than theoretically possible. Three novel adaptive schemes that increase the number of raw bits generated per photon were proposed and compared regarding the information rates they offer. Unlike in \cite{QKD:Zhou13, Kochman}, this work uses the singly occupied and singly empty bins to generate secret bits.

Like more informed modulation, high-dimensional QKD systems benefit from coding solutions that can utilize the temporal representation. Recall that in Section~\ref{section:time}, we described jitter errors, detector downtime, and dark counts. As a result, a successful channel code design should explicitly consider local and global channel properties.

In \cite{QKDlayered:Zhou13}, a layered scheme that partitions large-alphabet symbols into individual layers was presented. The key idea is to encode each bit layer using its own LDPC code, where the encoding can be done jointly or layer by layer. The joint channel was split into layer-by-layer channels based on the chain rule for mutual information. The work in \cite{QKDlayered:Zhou13} considered different channel models, incorporating both uniform and local errors. The LDPC codes considered in \cite{QKDlayered:Zhou13} were regular (unstructured) codes based on random constructions.

The practical feasibility of a multi-layered approach in the multi-dimensional setting was demonstrated in \cite{QKD:Zhong15} for the photon starved conditions associated with single photon detectors and long-distance propagation loss, as well as in \cite{Lee:19}, which showcased a successful implementation in both a laboratory setting and over deployed fiber.

Designed codes should be cognizant of the channel properties and have sufficient structure for fast decoding. Recent approaches \cite{QKD:Yang19, prisca} carefully construct LDPC codes optimized for the induced channel's local-global properties. The work in \cite{QKD:Yang19} was the first to consider finite-length code construction for this application; it proposed a balanced modulation scheme along with new construction of structured LDPC codes that, unlike the previous literature on LDPC codes for QKD, explicitly incorporates global error correction and local error correction in the Tanner graph of the code. Specifically, the Tanner graph of the code has a particular property that check nodes are organized into two disjoint types: the first type of (global) check nodes is connected to variable nodes at a group level, and the second type of (local) check nodes is connected only to variable nodes within each group.

Further refinement of code design for the combined  (quantized) Gaussian and uniform channels was done in \cite{prisca}. Here, spatially-coupled (SC) irregular repeat-accumulate
(IRA) codes strategically combine high-performance SC codes and IRA codes to overcome the dependencies in decoding amongst successive bits (and thus error propagation) present in the original multi-layer scheme \cite{QKDlayered:Zhou13} while being well matched to the induced channel. The improvements in the key rate are at least 20\% over the multi-layer scheme.

A rigorous treatment of systems with detector jitters is provided in \cite{QKD-C:boutrosS22}. This work computes the secret key rates possible with detector jitter errors and constructs codes for information reconciliation to approach these rates. In particular, the paper shows that even standard Reed-Solomon, BCH, and LDPC codes can achieve secret key rates much higher than the maximum achievable by polarization entanglement-based QKD.
 
 \section{Discussion and Open Questions}\label{section:discussion}
 
 In the preceding sections, we overviewed the fundamentals of QKD,  described time-entanglement QKD in detail, and discussed known channel coding methods for DV QKD. In addition to further exploration of ideas from information and coding theory based on the current results from \cite{QKD:Zhou13,QKD:KarimiSW20,QKDlayered:Zhou13, QKD:Yang19, prisca} for more realistic time-bin entanglement, we envision that the results summarized thus far can serve as the initial point of study and development of appropriate mathematical models (that are currently largely unavailable) for the following emerging quantum technologies.
 \begin{itemize} 
 \item Frequency combs. Due to their frequency scaling and long-term coherence, frequency combs offer a new, more robust platform for entangled photon generation \cite{frequency-comb}. 
\item Hyper-entanglement based QKD. In hyperentanglement \cite{hyper:Chang21}, information is represented on multiple bases. For example, the single basis of time-bin, as in the time-entanglement QKD is expanded to include polarization or angular momentum. 
\item Hybrid QKD schemes. Recent work \cite{djordjevic} considered a hybrid QKD protocol that simultaneously uses both CV and DV QKD. 
\item Quantum networks and conference key agreement. Quantum networks \cite{quantum-network} 
will provide secure multi-party communication provided the existance of an efficient conference key agreement multi-party protocol.
 \end{itemize}
 
Each of these technologies will individually benefit from the following:
 \begin{itemize} 
 \item Careful mathematical characterization of the appropriate channel models;
    \item Establishment of the capacity-style bounds and a rigorous analysis of considered mathematical models;
    \item Design of codes tailored to the specifics of the QKD channels;
    \item Design and implementation of low-latency decoding algorithms;
    \item Investigation of other types of codes beyond LDPC codes; and 
    \item Investigation of practical joint modulation and coding schemes.
\end{itemize}
 
 \ld{Additionally, identifying relevant attack models and providing security proofs for them is another fruitful direction for the high-dimensional time-bin QKD.}
 By relating open questions to the existing body of work in the classical setting, we hope to demonstrate that the barrier to entry into the quantum realm is not as high as it may seem and that there is an important role information and coding theory community can play in designing and developing quantum information systems of the future.
 
 \section*{Acknowledgement}
 This work was supported in part by NSF under grants FET 2007203 and  FET 2008728.
 We thank the following colleagues:
Murat Can Sar{\i}han for providing and explaining experimental data, and Esmaeil Karimi and Phil Whiting for general discussions on non-ideal detectors.
 
\bibliographystyle{ieeetr}
\bibliography{BITS-QKD-bib}

\begin{thebibliography}{10}

\bibitem{shannon}
C.~Shannon, ``Communication theory of secrecy systems,'' {\em Bell System
  Technical Journal}, vol.~28, pp.~656--714, Apr. 1949.

\bibitem{accenture}
Accenture, ``Untangling the future of quantum communications,'' {\em White
  Paper}, Dec. 2022.

\bibitem{Diamanti_2016}
E.~Diamanti, H.-K. Lo, B.~Qi, and Z.~Yuan, ``Practical challenges in quantum
  key distribution,'' {\em npj Quantum Information}, vol.~2, p.~16025, Nov.
  2016.

\bibitem{Shapiro}
Z.~Zhang, C.~Chen, Q.~Zhuang, F.~N. Wong, and J.~H. Shapiro, ``Experimental
  quantum key distribution at 1.3 gigabit-per-second secret-key rate over a 10
  d{B} loss channel,'' {\em Quantum Science and Technology}, vol.~3, p.~025007,
  Apr. 2018.

\bibitem{qinternet}
S.~Wehner, D.~Elkouss, and R.~Hanson, ``Quantum internet: A vision for the road
  ahead,'' {\em Science}, vol.~362, p.~eaam9288, Oct. 2018.

\bibitem{Lee:19}
C.~Lee, D.~Bunandar, Z.~Zhang, G.~R. Steinbrecher, P.~B. Dixon, F.~N.~C. Wong,
  J.~H. Shapiro, S.~A. Hamilton, and D.~Englund, ``Large-alphabet encoding for
  higher-rate quantum key distribution,'' {\em Optics Express}, vol.~27,
  pp.~17539--17549, Jun. 2019.

\bibitem{QKD:realistic}
F.~Xu, X.~Ma, Q.~Zhang, H.-K. Lo, and J.-W. Pan, ``Secure quantum key
  distribution with realistic devices,'' {\em Reviews of Modern Physics},
  p.~025002, May 2020.

\bibitem{QKD:BirnieCS22}
D.~{Birnie IV}, C.~Cheng, and E.~Soljanin, ``Information rates with non ideal
  photon detectors in time-entanglement based {QKD},'' {\em arXiv:2207.04146
  [cs.IT], {IEEE} Trans.\ Commun.\ to appear}, 2023.

\bibitem{Birnie:Thesis:2022}
D.~{Birnie IV}, ``{Costs of Detector Jitter in Time Entanglement Quantum Key
  Distribution},'' Master's thesis, Rutgers, The State University of New
  Jersey, School of Graduate Studies, 2021.

\bibitem{Cheng:Thesis:2022}
C.~Cheng, ``{The Effect of Detector Recovery-Time on Time-Entanglement QKD Key
  Rates and an Online Tool for Calculating Such Rates},'' Master's thesis,
  Rutgers, The State University of New Jersey, School of Graduate Studies,
  2022.

\bibitem{QKD:ekert91}
A.~K. Ekert, ``Quantum cryptography based on {B}ell’s theorem,'' {\em
  Physical Review Letters}, vol.~67, no.~6, pp.~661--663, 1991.

\bibitem{Kochman}
Y.~Kochman and G.~W. Wornell, ``On high-efficiency optical communication and
  key distribution,'' in {\em Proc. of the 2012 Information Theory and
  Applications (ITA)}, pp.~172--179, Feb. 2012.

\bibitem{QKD:Zhong15}
T.~Zhong, H.~Zhou, R.~D. Horansky, C.~Lee, V.~B. Verma, A.~E. Lita,
  A.~Restelli, J.~C. Bienfang, R.~P. Mirin, T.~Gerrits, S.~W. Nam, F.~Marsili,
  M.~D. Shaw, Z.~Zhang, L.~Wang, D.~Englund, G.~W. Wornell, J.~H. Shapiro, and
  F.~N.~C. Wong, ``Photon-efficient quantum key distribution using
  time–energy entanglement with high-dimensional encoding,'' {\em New Journal
  of Physics}, vol.~17, p.~022002, Feb. 2015.

\bibitem{QKD:wong21}
X.~Cheng, M.~C. Sarihan, K.-C. Chang, C.~Chen, F.~N.~C. Wong, and C.~W. Wong,
  ``Secure high dimensional quantum key distribution based on
  wavelength-multiplexed time-bin encoding,'' in {\em 2021 Conference on Lasers
  and Electro-Optics (CLEO)}, pp.~1--2, May 2021.

\bibitem{Maurer93a}
U.~Maurer, ``Secret key agreement by public discussion,'' {\em IEEE
  Transactions on Information Theory}, vol.~39, pp.~733--42, Mar. 1993.

\bibitem{cascade}
G.~Brassard and L.~Salvail, ``Secret key reconciliation by public discussion,''
  {\em Lecture Notes in Computer Science}, vol.~765, pp.~410--423, 1994.

\bibitem{books:GK2011}
A.~E. Gamal and Y.~Kim, {\em Network Information Theory}.
\newblock Cambridge University Press, 2011.

\bibitem{hyper:Chang21}
K.-C. Chang, X.~Cheng, M.~C. Sarihan, A.~K. Vinod, Y.~S. Lee, T.~Zhong, Y.-X.
  Gong, Z.~Xie, J.~H. Shapiro, F.~N. Wong, and C.~W. Wong, ``648
  {H}ilbert-space dimensionality in a biphoton frequency comb: entanglement of
  formation and {S}chmidt mode decomposition,'' {\em npj Quantum Information},
  vol.~7, p.~48, Mar. 2021.

\bibitem{QKD:KarimiSW20}
E.~Karimi, E.~Soljanin, and P.~Whiting, ``Increasing the raw key rate in
  energy-time entanglement based quantum key distribution,'' in {\em Proc. of
  the IEEE Asilomar Conference on Signals, Systems \& Computers}, Nov. 2020.

\bibitem{boutros}
D.~Elkouss, A.~Leverrier, R.~Alléaume, and J.~Boutros, ``Efficient
  reconciliation protocol for discrete-variable quantum key distribution,'' in
  {\em Proc. of the IEEE International Symposium on Information Theory (ISIT)},
  Jun. 2009.

\bibitem{girod}
D.~Varodayan, A.~Aaron, and B.~Girod, ``Rate-adaptive codes for distributed
  source coding,'' {\em Signal Processing}, vol.~86, pp.~3123--3130, Nov. 2006.

\bibitem{martin2010}
D.~Elkouss, J.~Martínez~Mateo, and V.~Martin, ``Information reconciliation for
  quantum key distribution,'' {\em Quantum Information \& Computation},
  vol.~11, p.~226–238, Mar. 2011.

\bibitem{Kasai2010InformationRF}
K.~Kasai, R.~Matsumoto, and K.~Sakaniwa, ``Information reconciliation for {QKD}
  with rate-compatible non-binary {LDPC} codes,'' in {\em 2010 International
  Symposium on Information Theory \& Its Applications}, pp.~922--927, Oct.
  2010.

\bibitem{polar}
P.~Jouguet and S.~Kunz-Jacques, ``High performance error correction for quantum
  key distribution using polar codes,'' {\em Quantum Information \&
  Computation}, vol.~14, p.~329–338, Apr. 2012.

\bibitem{kaist}
S.~Jeong, H.~Jung, and J.~Ha, ``Rate-compatible multi-edge type low-density
  parity-check code ensembles for continuous-variable quantum key distribution
  systems,'' {\em npj Quantum Information}, vol.~1, p.~6, Jan. 2022.

\bibitem{milicevic2018}
M.~Milicevic, C.~Feng, L.~Zhang, and P.~Gulak, ``Quasi-cyclic multi-edge {LDPC}
  codes for long-distance quantum cryptography,'' {\em npj Quantum
  Information}, vol.~4, p.~21, Dec. 2018.

\bibitem{QKD:Zhou13}
H.~{Zhou} and G.~{Wornell}, ``Adaptive pulse-position modulation for
  high-dimensional quantum key distribution,'' in {\em Proc. of the IEEE
  International Symposium on Information Theory (ISIT)}, Jul. 2013.

\bibitem{QKDlayered:Zhou13}
H.~Zhou, L.~Wang, and G.~Wornell, ``Layered schemes for large-alphabet secret
  key distribution,'' in {\em Proc. 2013 Information Theory and Applications
  Workshop (ITA)}, Feb. 2013.

\bibitem{QKD:Yang19}
S.~Yang, M.~C. Sarihan, K.-C. Chang, C.~W. Wong, and L.~Dolecek, ``Efficient
  information reconciliation for energy-time entanglement quantum key
  distribution,'' in {\em Proc. of the IEEE Asilomar Conference on Signals,
  Systems \& Computers}, Nov. 2019.

\bibitem{prisca}
S.~Yang, {\em Application-Driven Coding Techniques: From Cloud Storage to
  Quantum Communications}.
\newblock PhD thesis, University of California, Los Angeles (UCLA), 2021.

\bibitem{QKD-C:boutrosS22}
J.~J. Boutros and E.~Soljanin, ``Time-entanglement {QKD:} secret key rates and
  information reconciliation coding,'' {\em arXiv preprint arXiv:2301.00486},
  2023.

\bibitem{frequency-comb}
S.~K. Lee, N.~S. Han, T.~H. Yoon, and M.~Cho, ``Frequency comb single-photon
  interferometry,'' {\em Communications Physics}, vol.~1, p.~51, Sep. 2018.

\bibitem{djordjevic}
I.~B. Djordjevic, ``Hybrid {QKD} protocol outperforming both {DV}- and
  {CV}-{QKD} protocols,'' {\em IEEE Photonics Journal}, vol.~12, pp.~1--8, Feb.
  2020.

\bibitem{quantum-network}
M.~Proietti, J.~Ho, F.~Grasselli, P.~Barrow, M.~Malik, and A.~Fedrizzi,
  ``Experimental quantum conference key agreement,'' {\em Science Advances},
  vol.~7, p.~eabe0395, June 2021.

\end{thebibliography}

\begin{IEEEbiography}[{\includegraphics[width=1in,height=1.25in,clip,keepaspectratio]{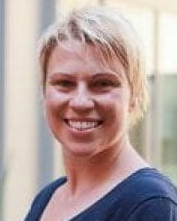}}]{Lara Dolecek}
is a professor of Electrical and Computer Engineering and (by courtesy) of Mathematics at the University of California, Los Angeles (UCLA). Prof. Dolecek earned a B.S., M.S., Ph.D. degrees in Electrical Engineering and Computer Sciences as well as an M.A. degree in Statistics from the University of California, Berkeley. She has served as an Associate Editor for IEEE Transactions on Information Theory, IEEE Transactions on Communications, and IEEE Communication Letters. She was a Secretary of the IEEE Information Theory Society, and is currently a Distinguished Lecturer of the society. Prof.~Dolecek received IBM Faculty Award (2014), Northrop Grumman Excellence in Teaching Award (2013), Intel Early Career Faculty Award (2013), University of California Faculty Development Award (2013), Okawa Research Grant (2013), NSF CAREER Award (2012), and Hellman Fellowship Award (2011). With her research group and collaborators, she received over dozen best paper awards. Her research interests span coding and information theory, graphical models, statistical methods, and algorithms, with applications to emerging systems for data storage and computing. 
 \end{IEEEbiography}

\begin{IEEEbiography}[{\includegraphics[width=1in,height=1.25in,clip,keepaspectratio]{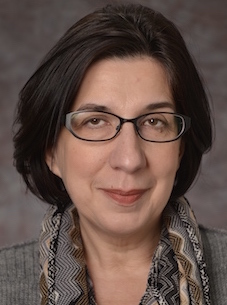}}]{Emina Soljanin} is a professor at Rutgers University. Before moving to Rutgers in 2016, she was a (Distinguished) Member of Technical Staff for 21 years in the Mathematical Sciences Research of Bell Labs.
Her interests and expertise are broad. Over the past quarter of the century, she has participated in numerous research and business projects, as diverse as power system optimization, magnetic recording, color space quantization, hybrid ARQ, network coding, data and network security, distributed systems performance analysis, and quantum information theory. She served as an Associate Editor for Coding Techniques, for the IEEE Transactions on Information Theory, on the Information Theory Society Board of Governors, and in various roles on other journal editorial boards and conference program committees.
 Prof.~Soljanin an IEEE Fellow, an outstanding alumnus of the Texas A\&M School of Engineering, the 2011 Padovani Lecturer, a 2016/17 Distinguished Lecturer, and 2019 President of the IEEE Information Theory Society.
 \end{IEEEbiography}
\end{document}